# EXCITATION OF BULK AND SURFACE WAKEFIELDS BY A RELATIVISTIC ELECTRON BUNCH IN PLASMA OF SEMICONDUCTORS AND SEMIMETALS

*V.A. Balakirev, I.N. Onishchenko*
*National Science Center "Kharkov Institute of Physics and Technology", Kharkiv, Ukraine*
*E-mail: onish@kipt.kharkov.ua*

The spatio-temporal structure of the bulk wakefields excited by a relativistic electron bunch in plasma of semiconductors and semimetals is studied. It is shown that these wakefield consists of the field of longitudinal plasmons and Cherenkov electromagnetic radiation, which is a set of eigen electromagnetic waves of the semiconductor or semimetal waveguide. A branch of the surface plasmons appears in a waveguide with an axial vacuum channel. The process of wake excitation of the surface plasmons by the relativistic electron bunch is also investigated. The intensity of the excited wake surface wave is determined.

## INTRODUCTION

Plasma phenomena in semiconductors and semimetals are an inherent property of these condensed media [1-7]. The density of electron-hole plasma in semiconductors with intrinsic conductivity (for example, silicon, germanium and other) is relatively low, on the order of $10^{10} \div 10^{13} \, cm^{-3}$, and is determined primarily by the width of the energy gap between the valence band and conduction band, and by the temperature of the material too [1,2]. In gapless semiconductors concentration of intrinsic carriers can be higher [8]. In doped semiconductors, the concentration of carriers (electrons in *n*-type semiconductors and holes in *p*-type semiconductors) can be significantly increased (within limits $10^{13} - 10^{18} \, cm^{-3}$) and it depends mainly on the dopant concentration [9].

An important characteristic of semiconductors and semimetals is also the frequency of the relaxation of carrier momentum [10,11]. There are two main mechanisms of carrier momentum relaxation in a crystal structure. This is the scattering of carriers by phonons of the crystal lattice and the scattering by impurities. Both mechanisms of momentum relaxation have a general nature and are associated with a violation of the periodicity of the crystal structure of a solid. With a decrease of the temperature of the solid, the frequency of collisions of carriers with phonons rapidly decreases, and scattering by impurity persists down to absolute zero.

Namely doped semiconductors could provide a sufficiently high concentration of carriers in solid-state plasma. But here a problem arises with the scattering of carriers by impurity. Doping of semiconductors, i.e. the introduction of impurities into a crystal, automatically causes an increase in the frequency of collisions of electron excitations with impurity atoms.

The situation is somewhat different with semimetals [12-14]. Unlike semiconductors, in semimetals (for example As, Sb, Bi and other) there is no energy gap between the valence band and the conduction band. In semimetals, these zones overlap weakly. The concentration of carriers in semimetals (electron-hole plasma) is within limits $10^{17} \div 10^{20} \, cm^{-3}$, that is higher than in semiconductors, but much lower than in metals $10^{22} \div 10^{23} \, cm^{-3}$. Therefore, in semimetals, as in conventional metals, a high carrier concentration is persisted, when the substance is cooled down to absolute zero. In pure semimetals, with decreasing temperature, the frequency of collisions of carriers with phonons, as in semiconductors, decreases. Then, the relaxation frequency of the carrier momentum reaches a certain constant level, determined by collisions of carriers with only of impurity atoms, which are always present in real crystals.

It is possible to talk about long-lived excitations in a solid-state plasma, as in any other medium, only if their eigen frequencies significantly exceed the frequencies of collisions with phonons and impurity atoms, $\omega \gg \nu_{col}$. Further we will assume that this condition is satisfied.

Since solid-state plasma is characterized by a high concentration of carriers and by a degree of homogeneity and stability, it seems very promising to use plasma of semiconductors and semimetals to realization wake methods of relativistic charged particles (primarily electrons and positrons) acceleration. In such scheme an intense relativistic electron bunch passes through vacuum channel of a semiconductor (semimetal) and excites wake eigen oscillations (plasmons, optical electromagnetic waves in the infrared range), which are in Cherenkov synchronism with a relativistic electron bunch $\omega = k(\omega)\beta_0 c$, $\omega$ is wave frequency, $k(\omega)$ is longitudinal wavenumber, $\beta_0 = v_0/c \approx 1$, $c$ is speed of light in vacuum. Excited wake waves can be used to accelerate charged particles [15,16].

We note that earlier in [17,18], the possibility of wake fields excitation by laser pulses or relativistic electron bunches in ion dielectrics was investigated. In [19-21] the possibility of excitation of the plasma branches of oscillations in semiconductors as a result of the beam-plasma instability development was studied. If there is a boundary in the waveguide, for example, an entrance end, then along with the wake electromagnetic field, the relativistic electron bunch excites transition electromagnetic radiation [22-26].

In the present work, the process of excitation of wake electromagnetic fields in the semiconductor (semimetal) waveguides by a relativistic electron bunch is investigated. Our aim is to obtain wake wave intensity, the frequency spectrum and a spatio-temporal structure of excited wakefields. Wake field includes the Langmuir wave of a solid-state plasma, and a set of eigen volume electromagnetic waves (volume polaritons) of the semiconductor (semimetal) waveguide.

To transport relativistic electron bunches, it is necessary to have a vacuum channel in the semiconductor (semimetal). On the other hand, in the presence of a medium-vacuum boundary in a waveguide, an additional branch of oscillations - surface plasmons can be excited [27-29]. In this work the process of wake surface plasmons excitation is studied. Surface plasmons are existed in the frequency range, where the permittivity is negative and there are no bulk waves in the medium.

## 1. STATEMENT OF THE PROBLEM

Let's consider the homogeneous semiconductor or semimetal cylinder of radius $b$, the side surface of which is covered with a perfectly conductive metal film. Below we will only talk about isotropic media. These are primarily crystals with a cubic lattice.

For the semiconductors in which impurity conductivity of one type dominates: donor or acceptor, the dielectric permittivity have the form

$$\varepsilon(\omega) = \varepsilon_{opt} - \frac{\omega_{pe,h}^2}{\omega(\omega + i\nu_{e,h})},$$

$\varepsilon_{opt}$ is optical permittivity, $\omega_{pe,h} = \sqrt{4\pi n_{e,h} e^2 / m_{e,h}}$ is plasma frequency of electrons and holes, $n_{e,h}$ − is free carrier concentration, $m_{e,h}$ is effective mass of electrons and holes, $\nu_{e,h}$ are effective collisions frequencies of electrons and holes.

In intrinsic semiconductors, as well as in semimetals, in which both electrons and holes are carriers, it is necessary to take into account their total contribution to the polarization of solid-state plasma. The dielectric constant for a solid-state electron-hole plasma has the form

$$\varepsilon(\omega) = \varepsilon_{opt} - \frac{\omega_{pe}^2}{\omega(\omega + i\nu_e)} - \frac{\omega_{ph}^2}{\omega(\omega + i\nu_h)}.$$

As noted above, we will consider oscillations and waves whose frequencies are significantly higher than the collision frequencies $\omega \gg \nu_{e,h}$. In this case, for the permittivity for all indicated substances, we obtain a simple expression

$$\varepsilon(\omega) = \varepsilon_{opt} - \frac{\omega_p^2}{\omega^2}, \qquad (1)$$

where $\omega_p^2 = \omega_{pe}^2 + \omega_{ph}^2$ is the plasma frequency in the most general case of electron - hole plasma.

The parameters $\varepsilon_{opt}$ and $\omega_p$ fully characterize the electrodynamic properties of the considered media. The values of these parameters for semiconductors and semimetals can differ greatly. Since in the future we will be interested only in the dielectric properties of these media, which are fully described by their permittivity $\varepsilon(\omega)$, then, for brevity, further we will consider semiconductors and semimetals similarly to dielectrics.

Along the axis of the dielectric waveguide, an axisymmetric REB moves uniformly and rectilinearly. The initial system of equations contains Maxwell's equations

$$rot\vec{E} = -\frac{1}{c}\frac{\partial \vec{H}}{\partial t}, \quad rot\vec{H} = \frac{1}{c}\frac{\partial \vec{D}}{\partial t} + \frac{4\pi}{c}\vec{j}_b$$

$$div\vec{D} = 4\pi\rho, \quad div\vec{H} = 0, \qquad (2)$$

$\rho_b, \vec{j}_b$ are charge density and current of an electron bunch, $\vec{D} = \hat{\varepsilon}\vec{E}$ is electric displacement field, $\hat{\varepsilon}$ is dielectric constant operator of a dielectric medium.

The system of Maxwell equations (2) describes the excitation of an electromagnetic field by external charges and currents in a condensed dielectric medium.

### 1.2. DETERMINATION OF THE GREEN FUNCTION

We will solve the problem of wake field excitation by an axisymmetric relativistic electron bunch in a dielectric waveguide as follows [18,25,26]. The problem will be solved as follows. We will find wakefield $\vec{E}_G$ (Green function) of elementary charge, having the form of a thin ring with charge $dQ$. Elementary charge density of an infinitely thin ring has the form

$$d\rho_b = \frac{dQ(r_0, t_0)}{v_0} \frac{\delta(r - r_0)}{2\pi r_o} \delta(t - t_0 - \frac{z}{v_0}), \qquad (3)$$

$$dQ = j_0(t_0, r_0) 2\pi r_0 dr_0 dt_0, \qquad (4)$$

$$j_0(r_0, t_0) = \frac{Q}{s_{eff} t_{eff}} R(r_0 / r_b) T(t_0 / t_b), \qquad (5)$$

where $Q$ is full charge of bunch, $t_0$ is time of entry of elementary charge, $r_0$ is radius ring, $v_0$ is bunch velocity, $t_b, r_b$ are characteristic duration and transverse bunch size, $R(r_0 / r_b)$ is function described transversal profile of bunch density, $s_{eff}$ is characteristic square of bunch transverse section,

$$s_{eff} = \pi r_b^2 \hat{\sigma}, \ \hat{\sigma} = 2\int_0^{b/r_b} R(\rho_0)\rho_0 d\rho_0.$$

The function $T(t_0 / t_b)$ describes longitudinal profile, $t_{eff}$ is effective bunch duration,

$$t_{eff} = \hat{\tau} t_b, \ \hat{\tau} = 2\int_0^\infty T(\tau_0) d\tau_0.$$

Let us represent the electromagnetic field excited by an elementary ring charge (3) as

$$\vec{E}_G(r, r_0, z, t - t_0) = dQ\vec{E}(r, r_0, z, t - t_0). \qquad (6)$$

Then the full electromagnetic field, excited by an electron bunch of finite dimensions, is found by summing (integrating) the fields of elementary ring bunches

$$\vec{E}(r, z, t) = \int_0^b 2\pi r_0 dr_0 \int_{-\infty}^t dt_0 j(r_0, t_0) \vec{E}(r, r_0, z, t - t_0). \qquad (7)$$

Taking into account relation (5), this expression can be written as follows

$$\vec{E}(r,z,t) = \frac{2\pi Q}{s_{eff} t_{eff}} \int_0^b R\left(\frac{r_0}{r_b}\right) r_0 dr_0 \times$$

$$\times \int_{-\infty}^t T\left(\frac{t_0}{t_b}\right) \vec{E}(r, r_0, z, t-t_0) dt_0. \quad (8)$$

The Green's function for the considered dielectric waveguide with a permittivity $\varepsilon(\omega)$ was obtained in [18] in the form of a series in Bessel functions

$$E_{Gz}(r, \bar{t}) = \frac{2i}{\pi b^2} dQ \sum_{n=1}^{\infty} \frac{J_0(\lambda_n r_0/b) J_0(\lambda_n r/b)}{J_1^2(\lambda_n)} S_n(\bar{t}), \quad (9)$$

where

$$S_n(\bar{t}) = \int_{-\infty}^{\infty} e^{-i\omega \bar{t}} \frac{d\omega}{\omega} \frac{k_\perp^2(\omega)}{\varepsilon(\omega) D_n(\omega)}, \quad (10)$$

$$D_n(\omega) = k_0^2 \varepsilon(\omega) - k_l^2 - \frac{\lambda_n^2}{b^2}, \quad (11)$$

$k_l = \omega/v_0$, $k_0 = \omega/c$, $\lambda_n$ are the roots of the Bessel function $J_0(x)$.

The zeros of the permittivity $\varepsilon(\omega) = 0$ are the poles of the integrand (10). Calculating the residues at the poles $\omega = \pm \omega_L - i0$, , где

$$\omega_L = \omega_p / \sqrt{\varepsilon_{opt}} \quad (12)$$

is Langmuir frequency of the solid-state plasma, we find the potential part of the Green's function

$$E_{Gz}^{(l)}(r, \bar{t}) = 2dQ \frac{k_L^2}{\varepsilon_{opt}} G(k_L r, k_L r_0) \vartheta(\bar{t}) \cos \omega_L \bar{t}, \quad (13)$$

$k_L = \omega_L / v_0$,

$$G(k_L r, k_L r_0) = \begin{cases} \frac{I_0(k_L r_0)}{I_0(k_L b)} \Delta_0(k_L r, k_L b), & r > r_0, \\ \frac{I_0(k_L r)}{I_0(k_L b)} \Delta_0(k_L r_0, k_L b), & r < r_0, \end{cases}$$

$\Delta_0(k_\alpha r, k_\alpha b) = I_0(k_\alpha b) K_0(k_\alpha r) - I_0(k_\alpha r) K_0(k_\alpha b)$.

The integrand has also poles

$$\omega = \pm \omega_{tn} - i0, \quad (14)$$

that are the roots of the equation

$$D_n(\omega) = \frac{\omega^2}{c^2} \varepsilon(\omega) - \frac{\omega^2}{v_0^2} - \frac{\lambda_n^2}{b^2} = 0. \quad (15)$$

Equation (15) determines the frequency spectrum of the eigen electromagnetic waves excited by the relativistic electron bunch in dielectric waveguide. From this equation we find the frequencies of electromagnetic waves

$$\omega_{tn} = \sqrt{\frac{\omega_n^2 + \omega_p^2 \beta_0^2}{\beta_0^2 \varepsilon_{opt} - 1}}, \quad \omega_n = \frac{\lambda_n v_0}{b}.$$

After calculation of the residues at the poles (14) of the integrand (10), we obtain the following expression for the electromagnetic part of the Green's function

$$E_{Gz}^{(t)} = \frac{4dQ}{b^2} \vartheta(\tau) \sum_{n=1}^{\infty} \frac{\omega_n^2}{\omega_n^2 + \omega_p^2 \beta_0^2} \frac{\Pi_n(r, r_0)}{\varepsilon_n} \cos \omega_{tn} \bar{t}, \quad (16)$$

where

$$\Pi_n(r, r_0) = \frac{J_0(\lambda_n r_0/b) J_0(\lambda_n r/b)}{J_1^2(\lambda_n)}, \quad \varepsilon_n = \varepsilon(\omega_{tn}).$$

Expression (16) describes bulk wake electromagnetic field excited by the infinitely thin electron ring bunch in dielectric waveguide. If $\omega_p = 0$ this expression for the wakefield coincides with that obtained in [15,30] for a dielectric waveguide in the absence of frequency dispersion of the dielectric constant $\varepsilon(\omega) = \varepsilon_0 = Const$.

Thus, we obtained the Green function, which describes the longitudinal component of the wake electric field excited by a ring relativistic electron bunch in the semiconductor waveguide. The Green function contains the longitudinal (potential) and electromagnetic (vortex) parts. The potential part is a field of longitudinal bulk plasmons. As for the electromagnetic part of the Green function, it contains a set of radial electromagnetic waves of semiconductor waveguide.

## 1.2. EXCITATION OF WAKEFIELDS BY AN ELECTRON BUNCH OF FINITE SIZE

The resulting electromagnetic field $\vec{E}(r, \tau)$ of the electron bunch (5) can be determined by summing the fields $\vec{E}_G$ of elementary electron ring charges. We first consider the excitation of wake plasmons by the relativistic electron bunch of finite dimensions.

For the wakefield of plasma oscillations we obtain the following expression

$$E_z^{(l)}(r, \tau) = E_L \Gamma_L(r) Z_\parallel(\omega_L \tau), \quad (17)$$

where

$$Z_\parallel(\omega \tau) = \frac{1}{t_{eff}} \int_{-\infty}^{\tau} T(\tau_0/t_b) \cos \omega(\tau - \tau_0) d\tau_0, \quad (18)$$

$$\Gamma_L(r) = \frac{2\pi}{s_{eff}} \int_0^b R(r_0/r_b) G(k_L r, k_L r_0) r_0 dr_0, \quad (19)$$

$$E_L = 2Q \frac{k_L^2}{\varepsilon_{opt}}. \quad (20)$$

The function $Z_\parallel(\omega \tau)$ describes the distribution of the wake field at a frequency $\omega$ in the longitudinal direction at each moment of time. We will consider an electron bunch with a symmetric longitudinal profile $T(\tau_0) = T(-\tau_0)$. The wake function $Z_\parallel(\omega \tau)$ is conveniently represented as in [18]

$$Z_\parallel(\omega \tau) = \frac{1}{\hat{\tau}} \left[ \hat{T}(\Omega) \vartheta(\tau) \cos \omega \tau - X(\bar{\tau}) \right], \quad (21)$$

where $\Omega = \omega t_b$, $\bar{\tau} = \tau/t_b$,

$$X(\bar{\tau}) = sign\tau \int_{|\bar{\tau}|}^{\infty} T(s) \cos \Omega(|\bar{\tau}| - s) ds,$$

$$\hat{T}(\Omega) = 2 \int_0^{\infty} T(s) \cos(\Omega s) ds, \quad s = t/t_b. \quad (22)$$

Behind a bunch $|\bar{\tau}| \gg 1$, the wake field (17) of plasma oscillations has the form of a monochromatic wave

$$E_z^{(l)}(r,\tau) = E_L \Gamma_L(r) \frac{\hat{T}(\Omega_L)}{\hat{\tau}} \cos\omega_L\tau, \ \Omega_L = \omega_L t_L. \quad (23)$$

We present the expressions for the Fourier amplitudes $\hat{T}(\Omega_L)$ for Gaussian longitudinal profiles of the electron bunch

$$T(\tau_0/t_b) = e^{-\tau_0^2/t_b^2}, \ \hat{T}(\Omega) = \sqrt{\pi} e^{-\Omega^2/4}, \ \hat{\tau} = \sqrt{\pi}. \quad (24)$$

Wake plasma wave is most efficiently radiated when the coherence condition is fulfilled $\omega_L t_b \leq 1$. If the inequality $\omega_L t_b \gg 1$ holds, then the electron bunch radiates incoherently and the amplitude of the wake plasma wave is exponentially small.

Let's consider an electron bunch with a Gaussian transverse profile

$$R(r) = e^{-r^2/r_b^2}. \quad (25)$$

When the condition $k_L b \gg 1$ is satisfied on the axis $r = 0$ the function $\Gamma_L(r)$ takes on the value

$$\Gamma_L(0) = -\frac{1}{2} e^{\rho_b} Ei(-\rho_b), \rho_b = \frac{k_L^2 r_b^2}{4}, \quad (26)$$

$$Ei(z) = \int_{-\infty}^{z} \frac{e^t}{t} dt$$

is integral exponential function. For thin $\rho_b \ll 1$ and wide $\rho_b \gg 1$ bunches the asymptotic representations for function (26) are

$$\Gamma_L(0) = \begin{cases} \frac{1}{2}\ln\left(\frac{1}{\rho_b}\right), & \rho_b \ll 1, \\ \frac{1}{\rho_b}, & \rho_b \gg 1. \end{cases}$$

Thus, with the full coherence of the Cherenkov excitation of wake plasma wave $\omega_L t_b \leq 1$, $k_L r_b \leq 1$ the wakefield on the axis of the waveguide takes the maximum value

$$E_z^{(l)}(0,\tau) = E_L \ln(2/k_L r_b) \cos\omega_L \tau. \quad (27)$$

Let us now consider the excitation of wake electromagnetic waves by an electron bunch in the dielectric waveguide. Using the electromagnetic Green function (16), we obtain the wake electromagnetic field as a superposition of radial modes

$$E_z^{(t)}(r,\tau) = \frac{4Q}{b^2}\sum_{n=1}^{\infty} L_n \Gamma_n \frac{J_0\left(\lambda_n \frac{r}{b}\right)}{J_1^2(\lambda_n)} Z_{\parallel}(\omega_m \tau), \quad (28)$$

$$\Gamma_n = \frac{2\pi}{s_{eff}}\int_0^b R\left(\frac{r_0}{r_b}\right) J_0\left(\lambda_n \frac{r}{b}\right) r_0 dr_0 =$$

$$= \frac{2}{\hat{\sigma}} \int_0^{1/\eta_b} R(\rho_0) J_0(\lambda_n \eta_b \rho_0) \rho_0 d\rho_0, \ \rho_0 = \frac{r_0}{r_b}, \eta_b = \frac{r_b}{b},$$

$$L_n = \frac{\omega_n^2}{\omega_n^2 + \omega_p^2 \beta_0^2}\frac{1}{\varepsilon_n},$$

$\varepsilon_n \equiv \varepsilon(\omega_m)$. The function $Z_{\parallel}(\omega\tau)$ is defined by the formula (18). For a symmetric electron bunch in the "wave zone" $\omega_m \tau \gg 1$, where the quasi-static field of the electron bunch is small, the wake field (28) is a superposition of radial monochromatic modes of the dielectric waveguide

$$E_z^{(t)}(r,\tau) = \frac{4Q}{b^2}\sum_{n=1}^{\infty} L_n \Gamma_n \frac{\hat{T}_n}{\hat{\tau}} \frac{J_0\left(\lambda_n \frac{r}{b}\right)}{J_1^2(\lambda_n)} \cos\omega_m \tau, \quad (29)$$

where $\hat{T}_n \equiv \hat{T}(\Omega_m)$ is Fourier component (22) of the function $T(s)$ at the dimensionless frequency $\Omega_m = \omega_m t_b$.

We also present an expression for the power of the wake electromagnetic radiation, which we define as the component of the total Poiting vector along the dielectric waveguide axis

$$P = \frac{c}{4\pi}\int_0^b \langle E_r^{(t)} H_\varphi^{(t)} \rangle 2\pi r dr.$$

Angle brackets mean averaging over high-frequency wakefield oscillations. As a result, for the radiated power we obtain the following expression

$$P = 2\beta_0 c \frac{Q^2}{b^2} \sum_{n=1}^{\infty} \frac{\omega_m^2 b^2}{v_0^2} L_n^2 \Gamma_n^2 \frac{\hat{T}_n^2}{\hat{\tau}^2} \frac{\varepsilon_n}{\lambda_n^2 J_1^2(\lambda_n)}. \quad (30)$$

For an electron bunch with a Gaussian longitudinal (27) and transverse (28) profiles, the coefficients $\Gamma_n$ and $\hat{T}_n$, which are determined by the specific form of the transverse and longitudinal density profiles of the bunch, have the form

$$\hat{T}_n = \sqrt{\pi}\exp\left(-\frac{1}{4}\Omega_m^2\right), \quad (31)$$

$$\Gamma_n = 2\int_0^{1/\eta_b} J_0(\lambda_n \eta_b \rho_0) e^{-\rho_0^2} \rho_0 d\rho_0.$$

When the condition $\eta_b \ll 1$ is satisfied, the expression for the coefficient $\Gamma_n$ is simplified

$$\Gamma_n = \exp\left(-\frac{1}{4}\lambda_n^2 \eta_b^2\right). \quad (32)$$

Accordingly, expression (29) for a Gaussian bunch, taking into account relations (31), (32), takes the form

$$E_z^{(t)}(r,\tau) = \frac{4Q}{b^2}\sum_{n=1}^{\infty} L_n e^{-\frac{1}{4}(\Omega_m^2 + \lambda_n^2 \eta_b^2)} \frac{J_0\left(\lambda_n \frac{r}{b}\right)}{J_1^2(\lambda_n)} \cos\omega_m \tau. \quad (33)$$

For the radiated power (30), we have

$$P = 2\beta_0 c \frac{Q^2}{b^2} \sum_{n=1}^{\infty} \frac{\omega_m^2 b^2}{v_0^2} L_n^2 e^{-\frac{1}{2}(\Omega_m^2 + \lambda_n^2 \eta_b^2)} \frac{\varepsilon_n}{\lambda_n^2 J_1^2(\lambda_n)}.$$

From expression (33) it follows that an electron bunch excites finite number of radial modes of electromagnetic radiation, for which the coherence condition $\omega_m^2 t_b^2 \leq 1$, $\lambda_n^2 r_b^2/b^2 \leq 1$ for excitation by an electron bunch is satisfied.

## 2. EXCITATION OF SURFACE PLASMONS

If there is a vacuum channel in a semiconductor or semimetal for transportation of the electron bunches, a qualitative change in the dispersion properties of a solid-state waveguide occurs. A new branch of oscillations appears - surface plasmons [26-28]. It is of interest to consider the possibility of the excitation of wakefield surface plasma waves by relativistic electron bunches in solid-state waveguides.

The waveguide system is a tube of outer radius $b$ made of semiconductor or semi-metallic material. The inner region $r < a$ is a vacuum cylindrical cavity. The region $b > r > a$ is filled with a homogeneous dielectric. The outer surface of the dielectric tube is covered with a perfectly conductive metal film that completely shields the electromagnetic field in the dielectric waveguide. A relativistic electron bunch moves in a vacuum channel.

## 2.1. DISPERSION PROPERTIES OF SURFACE WAVES OF SEMICONDUCTOR AND SEMIMETAL WAVEGUIDES

Let us dwell briefly on the dispersion properties of surface waves in a dielectric waveguide with the permittivity (1). Surface waves of the considered waveguide are described by the equation

$$\varepsilon(\omega) = -\frac{1}{q_v a}\frac{I_1(q_v a)}{I_0(q_v a)} q_d a \frac{\Delta_0(q_d a, q_d b)}{\Delta_1(q_d a, q_d b)}, \quad (34)$$

$$q_v = \sqrt{k^2 - k_0^2}, \quad q_d = \sqrt{k^2 - k_0^2 \varepsilon(\omega)}, \quad k_0 = \omega/c,$$

where $k$ is longitudinal wavenumber. First of all, note that on the plane ($\omega, k$) the region of existence of slow surface waves is bounded by the inequalities

$$\omega < kc, \quad \omega_L > \omega > 0. \quad (35)$$

In this region, on one side $q_v^2 > 0$, $q_d^2 > 0$, which ensures the surface nature of the waves in the vacuum region and in the dielectric medium, and on the other side, the permittivity is negative $\varepsilon(\omega) < 0$. In Fig. 1, this region is designated as $S_1$. In the regions $F_{1,2}$ to the left of the "light line" $\omega = kc$, the phase velocity of waves in vacuum exceeds the speed of light. Therefore, only fast eigen waves of a dielectric waveguide can exist in this region. For example, these are the waves that propagate in a vacuum channel and experience full internal reflection from the vacuum – medium boundary ($q_d^2 > 0$, $q_v^2 < 0$, region $F_1$). These waves can exist in the same frequency range as the surface waves. Also fast bulk electromagnetic waves can propagate in the waveguide in the frequency range $\omega > \omega_L$ (region $F_2$). Besides There are bulk electromagnetic waves in the region $S_2$, which are bulk ($q_d^2 < 0$) in the dielectric region and ones are surface ($q_v^2 > 0$) in vacuum range. Cherenkov excitation these waves was investigated in the previous section. And finally, in region $N$ eigen waves are absence.

The dispersion equation can be significantly simplified in the limiting case of a large radius of the vacuum channel $q_{v,d} a \gg 1$. This limiting case corresponds to the transition from a ring waveguide to a flat dielectric layer. The dispersion equation for a flat layer takes the form [22]

$$\varepsilon(\omega) = -\frac{q_d}{q_v}\tanh(q_d L), \quad (36)$$

where $L = b - a$ is the layer thickness.

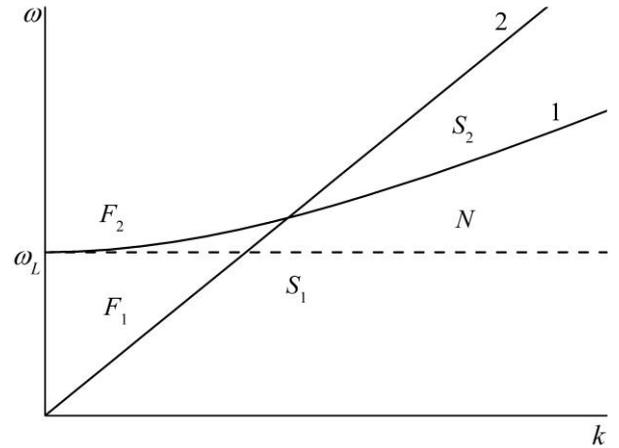

*Fig.1. Regions of existence of the waves of various types in a semiconductor (semimetal) waveguide, curve 1 is the dispersion dependence $\omega(k)$ of bulk electromagnetic waves in an infinite semiconductor medium, straight line 2 is the light line $\omega = kc$.*

For a thick layer $q_d L \gg 1$, dispersion equation (36) takes the form

$$\varepsilon(\omega) = -\frac{q_d}{q_v} \quad (37)$$

and describes the surface waves of a dielectric (semiconductor or semimetal) half-space. When $k \to \infty$ dispersion equations (36), (37) go over to the equation

$$\varepsilon = -1,$$

which determines the limiting frequency

$$\omega_\infty = \omega_L \sqrt{\frac{\varepsilon_{opt}}{\varepsilon_{opt}+1}}, \quad \omega_\infty < \omega_L. \quad (38)$$

For any thickness of the dielectric layer at $k \to \infty$ dispersion curve $\omega(k)$ tends to the cutoff frequency $\omega_\infty$.

Let us now consider the behavior of the dispersion curve $\omega(k)$ in the vicinity of the left boundary of the region of existence of surface waves (see Fig. 1). In this vicinity, the phase velocities of surface waves are close to the light speed in vacuum; therefore, these waves are of the greatest interest from the point of view of wake excitation of them by relativistic electron bunches. In this limiting case, dispersion equation (36) takes the form

$$\varepsilon(\omega) = -\frac{k_0\sqrt{1-\varepsilon(\omega)}}{q_v}\tanh\left(k_0 L\sqrt{1-\varepsilon(\omega)}\right). \quad (39)$$

It follows from this equation that at $q_v \to 0$ the frequency $\omega \to 0$, i.e. for any thickness of the dielectric layer, the dispersion curve $\omega(k)$ starts from the origin of coordinates $\omega = 0, k = 0$.

Let us now consider the features of the dispersion of surface waves in a tubular dielectric waveguide. At $k \to \infty$, the dispersion curve, as in the case of a flat layer, always reaches the limiting frequency (38). As for the behavior of the dispersion curve near the left boundary of the region of existence of a surface waves

$\omega = kc$, it is radically different from the case of a flat layer. Near this boundary, the condition

$$q_v a \ll 1 \quad (40)$$

is satisfied. This condition makes it possible to significantly simplify dispersion equation (34) and represent it in the form

$$\varepsilon(\omega) = -\frac{q_d a}{2} \frac{\Delta_0(q_d a, q_d b)}{\Delta_1(q_d a, q_d b)}. \quad (41)$$

At the boundary of the existence of surface waves $k = \omega/c$, the value $q_d$ can be represented as

$$q_d = \frac{1}{c}\sqrt{\omega_L^2 - \omega^2(\varepsilon_{opt} - 1)}.$$

Together with Eq. (41), we have the transcendental equation for determination of the initial frequency $\omega_{in}$ of the surface wave, from which the dispersion curve begins. In the frequency range $\omega_L > \omega > 0$, where surface waves exist, the value $q_d$ varies within finite limits

$$\kappa_L\sqrt{\varepsilon_{opt}} > q_d > \kappa_L,$$

where $\kappa_L = \omega_L/c$. And since the right-hand side of Eq. (41) as a function $q_d$, can vary over a wide range from 0 to, $-\infty$, then the value of the frequency $\omega_{in}$ at which the dispersion curve of surface waves begins changes over the entire interval of their existence $\omega_L > \omega_{in} > 0$. Recall that the longitudinal wavenumber $k_{in} = \omega_{in}/c$ corresponds to each value of the initial frequency $\omega_{in}$.

If, along with condition (40), the condition

$$q_d a \gg 1, \quad (42)$$

is satisfied, then instead of equation (44) in the vicinity of the boundary $\omega = kc$ we obtain

$$\varepsilon(\omega) = -\frac{k_0 a \sqrt{1-\varepsilon(\omega)}}{2}\tanh\left(k_0 L \sqrt{1-\varepsilon(\omega)}\right). \quad (43)$$

This equation differs significantly from the analogous equation (39) for a flat layer. In the limiting case of a thick cylindrical layer $q_d L \gg 1$, equation (43) for determination the initial frequency is simplified and takes the form

$$\frac{\varepsilon(\omega)}{\sqrt{1-\varepsilon(\omega)}} = -\frac{k_0 a}{2}.$$

Since the right side of this equation is large, from this equation we find the approximate value of the frequency at which the dispersion curve begins

$$\omega_{in} = \omega_L \sqrt{\frac{2\sqrt{\varepsilon_{opt}}}{\kappa_L a}}. \quad (44)$$

The condition for the applicability of this formula has the form

$$\kappa_L a / 2\sqrt{\varepsilon_{opt}} \gg 1.$$

Thus, in the case of a thick cylindrical layer, the dispersion curve of the surface wave on the boundary straight line $\omega = kc$ starts from the frequency (44), which is much lower than the frequency of volume plasmons of the semiconductor or semimetal.

Let us consider now the limiting case of a thin cylindrical layer

$$q_d L \ll 1. \quad (45)$$

In this case, the equation (43) for determination of the initial frequency $\omega_{in}$ is simplified and takes the form

$$\frac{\varepsilon(\omega)}{1-\varepsilon(\omega)} = -\frac{k_0^2 a L}{2}. \quad (46)$$

The right side of this equation is the product of the large $k_0 a \gg 1$ and small $k_0 L \ll 1$ parameters. Therefore, it can take, generally speaking, an arbitrary value. Equation (46) is equivalent to the following

$$\frac{\omega^2 - \omega_L^2}{\omega^2 - \frac{\omega_L^2}{\epsilon_0}} = \alpha_L \epsilon_0 \frac{\omega^2}{\omega_L^2}, \quad (47)$$

where $\alpha_L = \frac{\kappa_L^2 a L}{2}$ $\epsilon_0 = \frac{\varepsilon_{opt} - 1}{\varepsilon_{opt}} < 1$.

The root of this equation belonging to the region of existence of surface waves

$$\omega_L > \omega > 0 \quad (48)$$

has the form

$$\omega_{in}^2 = \frac{2\omega_L^2}{1 + \alpha_L + \sqrt{(1+\alpha_L)^2 - 4\alpha_L \epsilon_0}}. \quad (49)$$

We note that condition (48) is satisfied for all values of the parameter $\alpha_L$. If the dielectric cylinder layer is so thin that, along with condition (42), the condition $\alpha_L = \kappa_L^2 a L / 2 \ll 1$ is also satisfied, then the initial frequency is close to the frequency of volume plasmons

$$\omega_{in} = \omega_L \left(1 - \frac{\kappa_L a L}{4\varepsilon_{opt}}\right).$$

In the case of a sufficiently thick layer, when conditions (45) and $\alpha_L \gg 1$ are simultaneously satisfied, the value of initial frequency is significantly lower than the plasma frequency

$$\omega_{in} = \frac{\omega_L}{\sqrt{\alpha_L}} \ll \omega_L.$$

Finally, consider the limiting case of the vacuum channel of a small radius

$$q_d a \ll 1, \ q_d b \geq 1. \quad (50)$$

In this limiting case, dispersion equation (41) takes the form

$$\varepsilon(\omega) = -\frac{q_d^2 a^2}{2} \ln\left(\frac{1}{\nu q_d a}\right), \quad (51)$$

$\nu$ is Euler's constant. Solving this equation by the method of successive approximations, we find the value of the initial frequency

$$\omega_{in} = \omega_L \left(1 - \frac{\kappa_L^2 a^2}{4\varepsilon_{opt}} \ln\frac{1}{\nu \kappa_L a}\right).$$

The initial frequency at which the dispersion curve begins is slightly below the frequency of bulk plasmons. In this case the initial longitudinal wavenumber is $k_{in} = \omega_{in}/c$. If a more tougher condition is performed compared to (50), namely $q_d a \ll 1, q_d b \ll 1$, then instead of equation (41) we have

$$\varepsilon(\omega) = -\frac{q_m^2 a^2}{2}\ln\left(\frac{b}{a}\right).$$

Accordingly, for the initial frequency, we obtain

$$\omega_{in} = \omega_L\left[1 - \frac{\kappa_L^2 a^2}{4\varepsilon_{opt}}\ln\left(\frac{b}{a}\right)\right].$$

The initial frequency at which the dispersion curve begins is also slightly below the frequency of bulk plasmons.

## 2.2. GREEN FUNCTION FOR THE TUBULAR SEMICONDUCTOR WAVEGUIDES

The Green's function for a dielectric waveguide of the considered geometry was determined in [18]. Let's split the waveguide into three regions. Two regions are in the vacuum channel. One of them $v1$ is located inside the ring bunch ($r < r_0 < a$), and the second region $v2$ is located between the ring bunch and the inner boundary of the dielectric tube ($a > r > r_0$). The third region $d$ is a dielectric tube ($b > r > a$).

Expressions for the components of the electromagnetic field in each of the regions of a dielectric waveguide excited by an elementary ring bunch with a charge density (3) have the form [18]

$$E_{Gz}^{(v1)} = \frac{i}{\pi}\frac{dQ}{a^2}\int_{-\infty}^{\infty} e^{-i\omega\bar{t}}\frac{d\omega}{\omega}\frac{I_0(\kappa_\nu r)}{I_0(\kappa_\nu a)}\left[\frac{I(\kappa_\nu r_0)}{I_0(\kappa_\nu a)}\frac{1}{D(\omega)} + \kappa_\nu^2 a^2 \Delta_0(\kappa_\nu r_0, \kappa_\nu a)\right], \quad (52)$$

$$H_{G\varphi}^{(v1)} = \frac{1}{\pi}\frac{dQ}{a^2}\int_{-\infty}^{\infty} e^{-i\omega\bar{t}}\frac{d\omega}{\kappa_\nu c}\frac{I_1(\kappa_\nu r)}{I_0(\kappa_\nu a)}\left[\frac{I_0(\kappa_\nu r_0)}{I_0(\kappa_\nu a)}\frac{1}{D(\omega)} + \kappa_\nu^2 a^2 \Delta_0(\kappa_\nu r_0, \kappa_\nu a)\right],$$

$$E_{Gz}^{(v2)}(r) = \frac{i}{\pi}\frac{dQ}{a^2}\int_{-\infty}^{\infty} e^{-i\omega\bar{t}}\frac{d\omega}{\omega}\frac{I_0(\kappa_\nu r_0)}{I_0(\kappa_\nu a)}\left[\frac{I_0(\kappa_\nu r)}{I_0(\kappa_\nu a)}\frac{1}{D(\omega)} + \kappa_\nu a \Delta_0(\kappa_\nu r, \kappa_\nu a)\right], \quad (53)$$

$$H_{G\varphi}^{(v2)}(r) = \frac{1}{\pi}\frac{dQ}{a^2}\int_{-\infty}^{\infty} e^{-i\omega\bar{t}}\frac{d\omega}{\kappa_\nu c}\frac{I_0(\kappa_\nu r_0)}{I_0(\kappa_\nu a)}\left[\frac{I_1(\kappa_\nu r)}{I_0(\kappa_\nu a)}\frac{1}{D(\omega)} - \kappa_\nu^2 a^2 \Delta_1(\kappa_\nu r, \kappa_\nu a)\right],$$

$$E_{Gz}^{(d)} = \frac{i}{\pi}\frac{dQ}{a^2}\int \frac{I_0(\kappa_\nu r_0)}{I_0(\kappa_\nu a)}\frac{\Delta_0(\kappa_d r, \kappa_d b)}{\Delta_0(\kappa_d a, \kappa_d b)}\frac{e^{-i\omega\bar{t}}d\omega}{\omega D(\omega)}, \quad (54)$$

$$H_{G\varphi}^{(d)} = -\frac{1}{\pi}\frac{dQ}{a^2}\int \frac{I_0(\kappa_\nu r_0)}{I_0(\kappa_\nu a)}\frac{\Delta_1(\kappa_d r, \kappa_d b)}{\Delta_0(\kappa_d a, \kappa_d b)}\frac{\varepsilon(\omega) e^{-i\omega\bar{t}}d\omega}{\kappa_d c D(\omega)},$$

where

$$\Delta_n(\kappa_d r, \kappa_d b) = I_0(\kappa_d b)K_n(\kappa_d r) - (-1)^n I_n(\kappa_d r)K_0(\kappa_d b),$$

$$\kappa_d = \sqrt{k_l^2 - k_0^2 \varepsilon(\omega)} \equiv k_l\sqrt{\gamma_0^{-2} + \beta_0^2[1-\varepsilon(\omega)]},$$

$$\kappa_\nu = \sqrt{k_l^2 - k_0^2} \equiv k_l/\gamma_0,$$

$$D(\omega) = \frac{1}{\kappa_\nu a}\frac{I_1(\kappa_\nu a)}{I_0(\kappa_\nu a)} + \frac{\varepsilon}{\kappa_d a}\frac{\Delta_1(\kappa_d a, \kappa_d b)}{\Delta_0(\kappa_d a, \kappa_d b)}. \quad (55)$$

The integrands in (52)-(54) have only simple poles $\omega = \pm\omega_s - i0$, which are the roots of the equation

$$D(\omega) = 0. \quad (56)$$

We note that the integral Fourier representations (52)-(54) for the wakefield are valid for all isotropic media (dielectrics, gas plasma, plasma of semiconductors, semimetals and metals), the polarization properties of which are described by the scalar dielectric constant $\varepsilon(\omega)$. The roots of equation (56) determine the frequencies of the eigen waves of the dielectric waveguide, synchronous with the relativistic electron bunch. Within the framework of the considered model, a dielectric waveguide has a formally infinite number of branches of bulk slow electromagnetic waves (bulk polaritons) and one surface electromagnetic wave (surface plasmon). Cherenkov excitation by a relativistic electron bunch of bulk polaritons was considered in the previous section for the model of a dielectric waveguide with full filling. The presence of a vacuum channel does not introduce qualitative features into the picture of excitation of bulk polaritons; therefore, below we will focus on investigation the physical picture of excitation of only surface plasmons. We note that there is no pole $\varepsilon(\omega) = 0$ in the integrands (52)-(54). Therefore, an relativistic electron bunch moving in a vacuum channel does not excite bulk longitudinal plasmons. A similar situation occurs with wake excitation of surface waves in a plasma waveguide [31]. Calculating the residues at the poles $\omega = \pm\omega_s - i0$, we find the expression for the longitudinal component of the electric field of the surface wave

$$E_{Gz}^{(v,s)}(r,\bar{t}) = \frac{2dQ}{a^2}\frac{I_0(\kappa_{vs}r_0)I_0(\kappa_{vs}r)}{\Lambda_s I_0^2(\kappa_{vs}a)}\vartheta(\tau)\cos\omega_s\bar{t}. \quad (57)$$

where $\kappa_{vs} \equiv \kappa_\nu(\omega_s)$, $\Lambda_s \equiv \Lambda(\omega_s)$,

$$\Lambda(\omega) = \omega^2\frac{\partial D(\omega^2)}{\partial \omega^2}. \quad (58)$$

Accordingly, the expression for the magnetic field of the surface wave has the form

$$H_{G\varphi}^{(v,s)} = -\frac{2dQ}{a^2}\beta_0\gamma_0\frac{1}{\Lambda_s}\frac{I_0(\kappa_{vs}r_0)I_1(\kappa_{vs}r)}{I_0^2(\kappa_{vs}a)}\vartheta(\tau)\sin\omega_s\bar{t}.$$

We also give an expression for these components of the electromagnetic field in the dielectric

$$E_{Gz}^{(d,s)} = \frac{2dQ}{a^2}\frac{1}{\Lambda_s}\frac{I_0(\kappa_{vs}r_0)}{I_0(\kappa_{vs}a)} \times$$

$$\times \frac{\Delta_0(\kappa_{ds}r, \kappa_{ds}b)}{\Delta_0(\kappa_{ds}a, \kappa_{ds}b)}\vartheta(\tau)\cos\omega_s\bar{t}, \quad (59)$$

$$H_{G\varphi}^{(d,s)} = \frac{2dQ}{a^2}\frac{1}{\Lambda_s}\frac{I_0(\kappa_{vs}r_0)}{I_0(\kappa_{vs}a)}\frac{\omega_s}{\kappa_{ds}c}\frac{\Delta_1(\kappa_{ds}r, \kappa_{ds}b)}{\Delta_0(\kappa_{ds}a, \kappa_s b)}\vartheta(\bar{t})\sin\omega\bar{t},$$

here $\kappa_{ds} = \kappa_d(\omega_s)$.

Thus, we have obtained the expressions that describe the wake surface wave excited by a ring electron bunch in an isotropic dielectric waveguide with an axial vacuum channel.

## 2.3. EXCITATION OF SERFACE WAVES BY AN ELECTRON BUNCH OF A FINITE SIZE

The field of a surface wave excited by a relativistic electron bunch of finite sizes is found by summing the fields of elementary ring charges and currents according to formula (8). As a result, we obtain the following expressions for the components of the electromagnetic field of the surface wave

$$E_z^{(v,s)} = \frac{2Q}{a^2}\frac{1}{\Lambda_s}\frac{I_0(\kappa_{vs}r)}{I_0(\kappa_{vs}a)}\Gamma_{s0}Z_\parallel(\omega_s\tau), \qquad (60)$$

$$H_{G\varphi}^{(v,s)} = -\frac{2Q}{a^2}\beta_0\gamma_0\frac{1}{\Lambda_s}\frac{I_1(\kappa_{vs}r)}{I_0(\kappa_{vs}a)}\Gamma_s Z_\perp(\omega_s\tau),$$

$$E_z^{(d,s)} = \frac{2Q}{a^2}\frac{1}{\Lambda_s}\frac{\Delta_0(\kappa_{ds}r,\kappa_{ds}b)}{\Delta_0(\kappa_{ds}a,\kappa_{ds}b)}\Gamma_s Z_\parallel(\omega_s\tau),$$

$$H_\varphi^{(d,s)} = \frac{2Q}{a^2}\frac{1}{\Lambda_s}\frac{\varepsilon(\omega_s)\omega_s}{\kappa_{ds}c}\frac{\Delta_1(\kappa_{ds}r,\kappa_{ds}b)}{\Delta_0(\kappa_{ds}a,\kappa_{ds}b)}\Gamma_s Z_\perp(\omega_s\tau),$$

where

$$Z_\perp(\omega_s\tau) = \frac{1}{t_{eff}}\int_{-\infty}^{\bar{\tau}} T(t_0/t_b)\sin\omega_s(\tau-t_0)d\tau_0$$

$$\Gamma_{s0} = \frac{2\pi}{s_{eff}}\frac{1}{I_0(\kappa_{vs}a)}\int_0^a R\left(\frac{r_0}{r_b}\right)I_0(\kappa_{vs}r_0)r_0 dr_0$$

is the coupling coefficient of a relativistic electron bunch with a surface wave. We note that in the most interesting case

$$\kappa_{vs}a \equiv \frac{\omega_s a}{v_0\gamma_0} \ll 1. \qquad (61)$$

the longitudinal component of the electric field of the surface wave in the vacuum channel is practically uniform over the channel cross section, and the coupling coefficient for any transverse bunch profile $\Gamma_s = 1$. In this limiting case, the expression for the wake surface field (60) behind the electron bunch takes the form

$$E_z^{(v,s)} = \frac{2Q}{a^2}\frac{1}{\Lambda_s}\frac{\hat{T}(\Omega_s)}{\hat{\tau}}\cos\omega_s\tau. \qquad (62)$$

Here $\Omega_s = \omega_s t_b$.

Let us investigate the expressions for the field components of the wake surface wave in a number of limiting cases. First of all, consider the limiting case of a vacuum channel of small radius, when, along with inequality (61), the condition

$$\kappa_d a \ll 1$$

is fulfilled. In this case, the spectrum equation (56) can be simplified and represented in the form

$$\varepsilon(\omega) = -\frac{\kappa_d^2 a^2}{2}\left[\ln\left(\frac{1}{\nu\kappa_d a}\right) - \frac{K_0(\kappa_d b)}{I_0(\kappa_d b)}\right].$$

The roots of this equation can be easily found by the method of successive approximations. As a result, we obtain the following expression for the square of the surface wave frequency

$$\omega_s^2 = \omega_L^2\left\{1 - \frac{k_L^2 a^2}{2\varepsilon_{opt}}\left[\ln\left(\frac{1}{\nu\kappa_L a}\right) - \frac{K_0(\kappa_L b)}{I_0(\kappa_L b)}\right]\right\}.$$

The structure of the surface wave field has the form

$$E_z^{(s)} = E_L\frac{\hat{T}(\Omega_L)}{\hat{\tau}}\ln\left(\frac{1}{\nu k_L a}\right)\begin{cases}1, & r \leq a, \\ \frac{\Delta_0(k_L r,k_L b)}{\Delta_0(k_L a,k_L b)}, & r \geq a,\end{cases}$$

where $E_L$ is amplitude bulk plasma wave (23). At $\kappa_L a \to 0$, the surface wave frequency approaches to the frequency of bulk plasmons $\omega_L$. Accordingly, the structure of the wake surface wave continuously transforms into the field structure (17) of bulk plasmons of a semiconductor or semimetal at $r_b \to 0$.

The above consideration is valid when the radius of the vacuum channel and, accordingly, the transverse size of the electron bunch is significantly less than the wavelength of bulk plasmons $a \ll \lambda_L/2\pi$, $\lambda_L = 2\pi c/\omega_L$. A more realistic case is when the dimensions of the vacuum channel and the electron bunch are of the same order of the excited surface wave length or exceed it. We assume that the condition

$$\kappa_d a = k_0 a\sqrt{1+\epsilon(\omega)} \gg 1 \qquad (63)$$

is satisfied, where $\epsilon(\omega) = -\varepsilon(\omega)$. Then, even more so $\kappa_d a \gg 1$. The bunch is strongly relativistic so that requirement (61) is still satisfied. Under these conditions, the spectrum equation (56) for determination of the frequency of the wake surface wave takes the form

$$\epsilon(\omega)\left(1+\frac{k_0^2 a^2}{8\gamma_0^2}\right) = \frac{\kappa_d a}{2}\tanh(\kappa_d L). \qquad (64)$$

In the case of a thick layer

$$\kappa_d L = k_0 L\sqrt{1+\epsilon(\omega)} \gg 1 \qquad (65)$$

equation (64) is simplified

$$\epsilon(\omega) = \frac{\kappa_d a}{2}\left(1-\frac{k_0^2 a^2}{8\gamma_0^2}\right).$$

If the inequality holds

$$\frac{k_0^2 a^2}{4\gamma_0^2} \ll 1,$$

then this equation is equivalent to the following

$$\frac{\epsilon^2(\omega)}{\epsilon(\omega)+1} = \frac{k_0^2 a^2}{4}. \qquad (66)$$

We will assume the following condition also is satisfied

$$\frac{k_0^2 a^2}{4} = \pi^2\left(\frac{a}{\lambda_s}\right)^2 \gg 1, \qquad (67)$$

where $\lambda_s$ is the length of the surface wave. It is obviously then the inequality takes place

$$\epsilon(\omega) \gg 1. \qquad (68)$$

Accordingly, equation (66) is simplified and takes the form

$$\epsilon(\omega) = \frac{k_0^2 a^2}{4}.$$

This equation is equivalent to the following one

$$\frac{\omega_L^2}{\omega^2}-1 = \frac{\mu_L^2}{4}\frac{\omega^2}{\omega_L^2},$$

where

$$\mu_L = \frac{\omega_L a}{c\sqrt{\varepsilon_{opt}}}. \qquad (69)$$

The positive root of this equation determines the frequency of the wake surface wave

$$\omega_s = \frac{\omega_L}{\mu_L}\sqrt{2\left(1+\sqrt{\mu_L^2+1}\right)}. \qquad (70)$$

For the surface wave frequency it is always necessary to satisfy the condition $\omega_s < \omega_L$. This requirement restricts the admissible values of the parameter $\mu_L$ by the inequality

$$\mu_L > \sqrt{2(1+\sqrt{2})} \approx 2.2.$$

In this case, it is necessary to satisfy the inequality (68). If $\mu_L \gg 1$, then instead of (70) we have

$$\omega_s = \omega_{in} = \omega_L\sqrt{\frac{2}{\mu_L}} \ll \omega_L.$$

where $\omega_{in}$ is the initial frequency, which is determined by formula (47).

For the Gaussian longitudinal profile (27), instead of (62), we obtain

$$E_z^{(s)} = E_s e^{-\frac{\omega_s^2 t_b^2}{4}}\cos\omega_s\tau,$$

where

$$E_s = E_0\frac{1}{2\Lambda_s}, \quad E_0 = \frac{4Q}{a^2}.$$

When the condition of coherent excitation of a wake wave by an electron bunch $\omega_s^2 t_b^2/4 < 1$ is satisfied, the value $E_s$ is the amplitude of the longitudinal component of the electric field of the surface wave in the vacuum channel.

When condition (67) is satisfied, the expression for parameter $\Lambda_s$ (58) has the form

$$\Lambda_s = \frac{1}{2}\frac{\frac{\omega_L^2}{\omega_s^2}-\frac{1}{2}}{\frac{\omega_L^2}{\omega_s^2}-1}.$$

Taking into account the expression for the frequency of the surface wave (70) for the amplitude of the surface wave $E_s$, we have

$$E_s = E_0 F(\mu_L),$$

where

$$F(\mu_L) = \frac{\mu_L^2 - 2\sqrt{\mu_L^2+1}}{\mu_L^2 - \sqrt{\mu_L^2+1}}.$$

If $\mu_L \gg 1$, approximately we have $F(\mu_L) \simeq 1$ and for the amplitude of the surface wave we obtain a simple expression

$$E_s = \frac{4Q}{a^2}.$$

The wake wave amplitude is determined only by the bunch charge and the radius of the vacuum channel.

As an example, we consider a semiconductor waveguide based on gallium arsenide $n-GaAs$, which has the following parameters: electron concentration $n_e = 2\cdot 10^{17}\,cm^{-3}$, effective mass of electrons is $m_e = 0.067 m_0$, $m_0$ is electron rest mass, $\varepsilon_{opt} = 12.53$, collision frequency $\nu_e = 10^{11}\,s^{-1}$, radius of the vacuum channel is $a = 150\,\mu m$. The Langmuir frequency $\omega_L = 2.73\cdot 10^{13}\,rad/s$ and parameter value $\mu_L = 3.85$ correspond to these parameters. In accordance with formula (70) the frequency of the surface wave is equal to $\omega_s = 0.58\omega_L$. Accordingly the permittivity $\varepsilon(\omega_s) = -24,7$ and the parameter $\kappa_d a = 34.5$, so that inequalities (63), (68) are certainly satisfied. As a result, for the amplitude of the wake surface wave, we obtain the following value

$$E_s = 1.7\cdot 10^{-3} N_0 \,(V/cm),$$

where $N_0$ is the number of particles in the bunch. In particular, for the bunch with a number of particles $N_0 = 10^{11}$, the amplitude of the wake surface wave $E_s = 170\,MV/cm$ which is slightly lower than the value of the electric field strength inside the atoms of the solids, the typical value of which is of the order of $300\,MV/cm$ [32]. The surface wave length is $\lambda_s = 119\,\mu m$.

Let us now consider the limiting case of a thin dielectric layer

$$\kappa_d L \ll 1. \qquad (71)$$

This inequality will be corrected below. In this limiting case, the spectrum equation (71) is simplified and takes the form

$$\varepsilon(\omega) = -\frac{1}{2}\kappa_d^2 aL.$$

Using an explicit expression for the dielectric constant (1), this equation can be transformed to the form

$$\frac{\omega^2 - \omega_L^2}{\omega^2 - \frac{\omega_L^2}{e_0}} = e_0 \alpha_L \frac{\omega^2}{\omega_L^2}, \qquad (72)$$

where

$$e_0 = \frac{\varepsilon_{opt} - \beta_0^{-2}}{\varepsilon_{opt}}, \quad \alpha_L = \frac{\kappa_L^2 aL}{2}.$$

The root of the biquadratic equation with respect to frequency (72), belonging to the frequency range (48) of surface waves, has the form

$$\omega_s^2 = \frac{2\omega_L^2}{1+\alpha_L + \sqrt{(1+\alpha_L)^2 - 4\alpha_L e_0}} \leq \omega_{in}^2, \qquad (73)$$

In the ultrarelativistic case $\beta_0 = 1$, we have $e_0 = \epsilon_0$ and the frequency (73) coincides with the initial frequency $\omega_{in}$ (49). If the cylindrical dielectric layer is so thin that the stronger condition is satisfied

$$\alpha_L \ll 1, \qquad (74)$$

in comparison with (71), then the expression for the frequency of the excited surface wave follows from the formula (73)

$$\omega_s^2 = \omega_L^2\left[1-(1-e_0)\alpha_L\right] = \omega_L^2\left(1-\frac{\alpha_L}{\beta_0^2 \varepsilon_{opt}}\right). \qquad (75)$$

In the considered case of a thin cylindrical layer, the expression for the longitudinal component of the electric field of the wake surface wave in the vacuum channel behind the electron bunch has form (62), where

$$\Lambda_s = \frac{1}{2}\left[1 + \frac{\alpha_L(1-e_0)w_s^4}{(1-w_s^2)^2}\right], \quad (76)$$

$$w_s^2 = \frac{2}{1+\alpha_L+\sqrt{(1+\alpha_L)^2-4\alpha_L e_0}} < 1.$$

For an electron bunch with a Gaussian longitudinal profile, the expression for the wakefield (62) taking into account the relation (76) takes the form

$$E_z^{(v,s)} = E_0\left[\frac{(1-w_s^2)^2}{(1-w_s^2)^2+\alpha_L(1-e_0)w_s^4}\right]e^{-\frac{\omega_s^2 t_b^2}{4}}\cos\omega_s\tau. \quad (77)$$

For a short bunch $\omega_L^2 t_b^2/4 \ll 1$ and a thin cylindrical layer (74) instead of (77) we obtain

$$E_z^{(v,s)} = 2Q\frac{k_L^2}{\varepsilon_{opt}}\frac{L}{a}\cos\omega_L\tau,$$

and the frequency of the surface wave (75) is close to the frequency of bulk plasmons.

## CONCLUSION

Solid-state plasma of semiconductors and semimetals is characterized by a high concentration of charged particles (electrons and holes), a degree of homogeneity and stability of solid-state plasma. That is why solid-state plasma is attractive from the point of view of using it as a medium for exciting intense wake waves by relativistic electron bunches.

Semiconductors and semimetals are considered as plasma-like media. In crystals of dielectrics of this group, there are two branches of oscillations in the infrared and terahertz frequency ranges. These are, first of all, longitudinal plasma oscillations of an electron-hole plasma. And also in the infrared range there is a branch corresponding to optical slow electromagnetic waves (transverse bulk polaritons) of the media. For all of these branches, analytical expressions for the wake electromagnetic field excited by a relativistic electron bunch are obtained and investigated. It is shown that in the infrared (microwave) frequency range, the excited wakefield electric field consists of a potential monochromatic wave belonging to the branch of longitudinal plasmons and a set of electromagnetic eigenwaves of a dielectric waveguide (transverse polaritons).

In the presence of a vacuum channel surface plasmons appear in the spectrum of eigen waves of the semiconductor (semimetal) waveguide. Surface plasmons exist in the frequency range $\omega_L > \omega > 0$. Note that, in this frequency range, there are no volume transverse polaritons. The process of excitation of wake surface plasmons by a relativistic electron bunch propagating in a vacuum channel is also investigated. It is shown that relativistic electron bunches of several tens microns in sizes can serve as an effective tool for coherent excitation of accelerating wakefields in the infrared range. Plasma of semiconductors and semimetals is of considerable interest for creating solid-state wake accelerators of charged particles.